# Topological Transitions in Metamaterials

Harish N S Krishnamoorthy[1,2, §], Zubin Jacob[3, §], Evgenii Narimanov[4], Ilona Kretzschmar[5], and

Vinod M. Menon[1,2,*]

[1]*Department of Physics, Queens College of the City University of New York (CUNY), Flushing, NY 11367*
[2] *Department of Physics, Graduate Center of the City University of New York (CUNY), New York, NY 10016*
[3]*Dept. of Electrical and Computer Engineering, University of Alberta, Edmonton, Canada, T6G 2V4*
[4]*Birck Nanotechnology Center, School of Electrical and Computer Engineering, Purdue University, West Lafayette, IN 47907*
[5]*Dept. of Chemical Engineering, City College of the City University of New York (CUNY) New York, NY 10031*

---

\* *Corresponding Authors: vmenon@qc.cuny.edu*
§ These authors contributed equally to this work

**The ideas of mathematical topology play an important role in many aspects of modern physics – from phase transitions to field theory to nonlinear dynamics (Nakahara M (2003) in *Geometry, Topology and Physics*, ed Brewer DF (IOP Publishing Ltd, Bristol and Philadelphia), Monastryskiy M (1987) in *Riemann Topology and Physics*, (Birkhauser Verlag AG)). An important example of this is the Lifshitz transition (Lifshitz IM (1960) Anomalies of electron characteristics of a metal in the high-pressure region, *Sov Phys JETP* 11: 1130-1135), where the transformation of the Fermi surface of a metal from a closed to an open geometry (due to e.g. external pressure) leads to a dramatic effect on the electron magneto-transport (Kosevich AM (2004) Topology and solid-state physics. *Low Temp Phys* 30: 97-118). Here, we present the optical equivalent of the Lifshitz transition in strongly anisotropic metamaterials. When one of the components of the dielectric permittivity tensor of such a composite changes sign, the corresponding iso-frequency surface transforms from an ellipsoid to a hyperboloid. Since the photonic density of states can be related to the volume enclosed by the iso-frequency surface, such a topological transition in a metamaterial leads to a dramatic change in the photonic density of states, with a resulting effect on every single physical parameter related to the metamaterial – from thermodynamic quantities such as its equilibrium electromagnetic energy to the nonlinear optical response to quantum-electrodynamic effects such as spontaneous emission. In the present paper, we demonstrate the modification of spontaneous light emission from quantum dots placed near the surface of the metamaterial undergoing the topological Lifshitz transition, and present the theoretical description of the effect.**

### Introduction

Metamaterials are artificially engineered structures in which the length scale of the substructures is much smaller than the wavelength of light. In such systems, it is the subwavelength features that control the macroscopic electromagnetic properties. Metamaterials have attracted much attention over the last decade owing to their wide applications ranging from subwavelength imaging (1-6), hyperlens (5,6) to optical cloaking (7-9) and control of spontaneous emission (10-13). Recent advances in nanofabrication technologies allow for realizing metamaterial systems where the dielectric permittivity and magnetic permeability tensor can be designed and engineered at will (7-9, 14). This ability to control the material parameters also allows us to mimic and study physical processes which are difficult, if not impossible, to study by any other methods (15-18).

Here, we present the optical equivalent of an Electronic Topological Transition (ETT), a phenomena which occurs when a closed Fermi surface transitions to an open one leading to drastic effects on the thermodynamic and kinetic properties of electrons in metals (19). We demonstrate that engineered metamaterials can show the optical equivalent of this effect – the optical topological transition (OTT) wherein the very nature of the electromagnetic radiation in the metamaterial undergoes a significant change. Effects on the kinetic properties such as the dynamics of propagating waves supported by the system to thermodynamic properties such as the electromagnetic energy density gets modified at the transition point. This modification can be probed by following the light-metamaterial interaction using a quantum emitter.

The optical isofrequency curve $\omega(\vec{k}) = const$ can be engineered by tailoring the dielectric tensor $\overleftrightarrow{\varepsilon}(\vec{r})$. Metal dielectric composites can make the permittivity anisotropic and can considerably distort the topology of the isofrequency curve. We consider the case of metal-dielectric composite metamaterials which have a uniaxial form of the dielectric tensor $\overleftrightarrow{\varepsilon}(\vec{r}) = diag(\varepsilon_{xx}, \varepsilon_{yy}, \varepsilon_{zz})$ where $\varepsilon_{xx} = \varepsilon_{yy} = \varepsilon_\parallel$ and $\varepsilon_{zz} = \varepsilon_\perp$. The isofrequency curve for the extraordinary (TM-polarized) waves propagating in such strongly anisotropic metamaterial is given by:

$$\frac{k_x^2 + k_y^2}{\varepsilon_\perp} + \frac{k_z^2}{\varepsilon_\parallel} = \frac{\omega^2}{c^2}$$

Closed isofrequency surfaces different from a simple sphere (eg: ellipsoidal) can occur in these metamaterials when $\varepsilon_\parallel > 0 \text{ and } \varepsilon_\perp > 0$. On the other hand an extreme case of iso-frequency curve modification occurs when the dielectric constants show opposite sign ($\varepsilon_\parallel < 0 \text{ and } \varepsilon_\perp > 0$) such that the iso-frequency curve opens up into a hyperboloid surface. We can design the metamaterial such that the metallic dispersion leads to an OTT in the isofrequency surface from ellipsoidal to hyperboloidal as the wavelength is changed. By controlling the relative fill fraction of the dielectric and metal we can tune the transition wavelength wherein one of the components of the dielectric permittivity tensor is close to zero ($\varepsilon_\perp > 0$ and $\varepsilon_\parallel \approx 0$).

Since the density of the photonic states in the metamaterial can be related to the volume enclosed by the corresponding surface of constant frequency (20), this topological transition from the closed (elliptical) constant frequency surface at $\varepsilon_\parallel > 0$ to an open (hyperbolic) surface of constant frequency at $\varepsilon_\parallel < 0$ (Fig.1) in the lossless effective medium limit is accompanied by a dramatic change in the density of states from a finite to an infinite value, resulting in a singularity in the corresponding thermodynamic functions of the system.

Such "Lifshitz transition" in optical metamaterials is characterized by the appearance of additional electromagnetic states in the metamaterial in the hyperbolic regime, which have wavevectors much larger than those allowed in vacuum. Light-matter interaction is enhanced due to the presence of hyperbolic metamaterial states, resulting in strong effect on related quantum-electrodynamical phenomena, such as the spontaneous emission. As the photonic density of states governs the spontaneous emission rate, signature of this transition can be probed by studying the modification of the radiative lifetime of a quantum emitter placed near the metamaterial structure.

**Results and Discussion**

The decay rate near a half space of a metamaterial for a dipole-like emitter is given by

$$\Gamma = \Gamma_{vac} + \Gamma_{plasmon} + \Gamma_{high-k}$$

where $\Gamma_{vac}$ is the decay rate due to propagating waves in vacuum, $\Gamma_{plasmon}$ is the enhanced decay due to plasmonic modes supported by the metal-dielectric composite metamaterial and $\Gamma_{high-k}$ is the decay rate enhancement due to the high wavevector states which appear only beyond the Lifshitz transition. In the near field of the metamaterial, when d << λ the decay rate is dominated by the contribution from the large wavevector spatial modes ($\Gamma_{high-k}$). This dominant near field contribution can be written as

$$\Gamma_{high-k} \approx \frac{\mu_\perp^2 \, \text{Im}(r_p)}{8\hbar d^3}$$

where $\mu_\perp$ is the dipole moment of the perpendicularly oriented dipole and $r_p$ is the plane wave reflection coefficient of p-polarized light (see *Supporting Information*). In the zero loss limit, we note the sharp increase in decay rate caused by these states which occur only in the hyperbolic dispersion regime. In a hyperbolic metamaterial half space where $real(\varepsilon_\parallel) < 0$

$$Im(r_p) = \sqrt{\frac{2|\varepsilon_\parallel|\varepsilon_\perp}{1 + |\varepsilon_\parallel|\varepsilon_\perp}}$$

and for $(real(\varepsilon_\parallel) > 0)$ we have elliptical dispersion with $\text{Im}(r_p) = 0$. We thus introduce the topological transition parameter $\alpha = \text{Im}(r_p)$ which is proportional to the local density of electromagnetic modes and thus characterizes the emergence of metamaterial states. Note that in the zero loss effective medium limit there is a discontinuity in the first derivative of $\alpha$ at the transition wavelength. In our experiment, we study the effect of this transition parameter on the spontaneous emission rate of a dipole emitter.

Any practical metamaterial realization has non-idealities emerging from the finite patterning scale that achieves the unique electromagnetic response and also from absorption losses due to metallic building blocks. We first study the effect of losses on the spontaneous emission rate of a dipole emitter placed near an effective medium metamaterial as the wavelength passes through the transition point. The effect of dispersive and lossy effective medium dielectric constants on the topological transition parameter (proportional to the spontaneous emission rate) is shown in Fig. 2a. While the introduction and further increase of the amount of losses in the dielectric permittivity reduces the transition to a smooth crossover, the change in the isofrequency surface topology still manifests itself in a greatly enhanced rate of the spontaneous emission. As a result, one can observe the effects of this topological transition even in relatively lossy metamaterials.

Another effect to consider is the finite size of the unit cell forming the metamaterial with the extremely anisotropic dielectric response achieved using alternating subwavelength layers of metal and dielectric. A deep subwavelength unit cell ($a << \lambda$) is simulated to mimic the effective medium response and shed light on the metamaterial states corresponding to the experimental situation. The effect of varying unit cell sizes on the spontaneous emission is shown in Fig. 2b. A strong change in the spontaneous emission rate is observed in this multilayer structure at the transition point, with the emergence of extra metamaterial states in this multilayer structure related to the coupled plasmons which occur only on one side of the transition (see *Supporting Information*).

To observe the signature of the predicted topological transition manifested through enhancement in spontaneous emission rate, we investigate a metamaterial structure that consist of alternating subwavelength layers of metal (silver) and dielectric (TiO$_2$), with multiple quantum dot emitters positioned on its top surface - see Fig. 3a. The dielectric constants of the constituent thin films are extracted using ellipsometry and the effective medium parameters are shown in Fig. 3b. This structure is designed to have $\varepsilon_\parallel \approx 0$ around the emission maximum of the quantum dots. The photoluminescence (PL) from the quantum dots have a full width at half maximum (FWHM) of ~ 40 nm which allows us to investigate the phase space of both ellipsoidal and hyperbolic dispersion regimes using the same sample.

To isolate the effects of the non-radiative decay due to metal, we also measured the spontaneous emission rates of quantum dots on a control sample that consists of a thin layer of the silver which has similar absorption losses as the metamaterial (as ascertained by ellipsometric measurements of the imaginary part of the dielectric constant) and hence has an approximately equal contribution to non-radiative lifetime decrease (as confirmed by simulations using the semi-classical model). Details of the control sample are discussed in the *Methods* section.

Time resolved photoluminescence (PL) measurements were carried out using a time correlated single photon counting technique (See *Methods* section for details). The time resolved PL data from the metamaterial sample, the control sample and the glass substrate are shown Fig 4(a). These measurements were carried out at a wavelength of 621 nm, which is very close to the topological transition wavelength. The large change in the spontaneous emission lifetime of the quantum dots between those on the metamaterial and the glass substrate is primarily due to the excitation of the metamaterial states. In addition, the difference in reflectivity and non-radiative decay due to the metal component also play a role in the decrease of the lifetime. This interpretation is confirmed on comparing the lifetime with a control sample which takes into account the non-radiative decay and the additional plasmonic decay route. The spontaneous emission rate shows an enhancement by a factor of ~ 3 in the metamaterial when compared to the control sample which is attributed solely due to the metamaterial states. **The overall reduction in the lifetime of the quantum dots when compared to those on a glass substrate is ~ 11**.

In Fig 4b, we show the change in lifetime as a function of wavelength for the quantum dots on the bare substrate, on the control sample and the metamaterial sample. It may be noticed that lifetime increases with wavelength on both the glass substrate and control samples. This is due to the size distribution of quantum dots and the weak dependence of the oscillator strength on the energy (21). On the contrary, the metamaterial sample shows a decrease in the lifetime as a function of wavelength as well as the shortest lifetime due to the presence of the metamaterial states. The decrease in the lifetime on the control sample is related to the well-known plasmonic decay route. The coupling of the emission from the quantum dots into the metamaterial states is also verified using steady state PL measurements in reflection geometry. The coupling to the *high-k* states is evident through the reduction in the PL intensity emitted in the direction away from the metamaterial sample (see *Supporting Information*).

We now present our experimental studies of the transition. First, we look at the radiative lifetimes close to the transition point. To make any conclusions regarding the effect of the optical topological transition on the radiative lifetimes from the measurement of the actual decay rates, the contribution of the metamaterial states to the total decrease in lifetime has to be distinguished from non-radiative decay. It is also critical to distinguish the role of plasmons in a single unit cell of the metamaterial (control sample) from the unique metamaterial states which cause the transition that we are studying here. Furthermore, the size inhomogeneity (~5%) in the ensemble of quantum dots affects their lifetime due to varying oscillator strengths (21). To account for all other effects that alter the lifetime of the quantum dots, we look at the lifetime of dots on the metamaterial normalized by the lifetime near the control sample as shown in Fig. 5. While the combination of metal losses and finite thickness of layers leads to a smooth crossover, the signature of the transition is clear from the reduction in the normalized lifetime in the hyperbolic regime.

To demonstrate the effect of the optical topological transition on the light-metamaterial interactions, we compare the lifetimes at the onset of the hyperbolic regime to those deep in the elliptical "phase". As the choice of our quantum emitter limits the measurement wavelength to that within the quantum dot emission linewidth, to access the elliptical phase we fabricate another sample with a different volume ratio of silver and TiO$_2$. With the thicknesses of each silver and TiO$_2$ layers of 8 nm and 25 nm respectively (more details on the sample preparation and the measurements can be found in the *Supporting Information*), the dielectric permittivity of the system was determined to be 0.43 in the direction parallel to the layers and 6.35 in the direction perpendicular to the layers. The "elliptical" sample, measured at the same wavelength of 612 nm as the "hyperbolic" shows substantially longer normalized lifetime of 0.48 as compared to 0.09 in hyperbolic sample (see the *Supporting Information* figures S3, S4 and S5 for the details of the measurements). This is consistent with the theoretical prediction for the normalized radiation emission lifetimes in these two systems ($\tau_{elliptical} \sim 0.39$ and $\tau_{hyperbolic} \sim 0.06$). Thus, the changes in the lifetime observed experimentally between the elliptic and the hyperbolic metamaterial structures can be attributed to the increase in the photonic density of states that manifests when the system goes through the topological transition in its iso-frequency surface.

In summary, we have shown that metamaterials can control the topology of the iso-frequency surface leading to an OTT, akin to a Lifshitz transition of electrons in metals. The consequence of the transition is the appearance of a large number of new electromagnetic states supported by the metamaterial. We have established the occurrence of OTT in a metamaterial designed to exhibit this transition near the emission wavelength of a quantum emitter by demonstrating reduction in its lifetime due to the presence of excess states. As mentioned, a host of interesting effects can transpire at the transition wavelength such as the sudden appearance of resonance cones, which are characteristic of hyperbolic metamaterials (22), enhanced nonlinear effects as well as abrupt changes in the electromagnetic energy density. We expect the OTT to be the basis for a number of applications of both fundamental and technological importance using metamaterial- based control of light-matter interaction.

## Materials and Methods

### Structure Details

The metamaterial structure studied in the present work was fabricated using physical vapor deposition. The structure consisted of alternating layers of silver and titanium dioxide ($TiO_2$) with thicknesses of 9 nm and 22 nm, respectively. Silver layers were deposited by evaporating silver pellets (obtained from Kurt J Lesker) at a pressure of $3 \times 10^{-6}$ mbar. TiO2 layers deposition were deposited by evaporating Titanium wires (obtained from Alfa-Aesar) in the presence of oxygen at a pressure of $2 \times 10^{-3}$ mbar. The particular structure investigated here consists of five periods of the structure on a 2.54 $cm^2$ glass cover slip substrate with the top $TiO_2$ layer also acting as a spacer. The control sample consists of a single period of silver and $TiO_2$ of thicknesses 9 nm and 22 nm, respectively. This structure takes into account the plasmon induced (radiative and non-radiative) modification in the lifetime present in the metamaterial structure. A 25% v/v solution of colloidal CdSe/ZnS core-shell quantum dots in toluene (obtained from Evident Technologies) was deposited on the metamaterial and control structure via spin coating.

### Photoluminescence

Time resolved photoluminescence (PL) measurements were carried out using time correlated single photon counting (TCSPC) technique with time resolution $\geq 200$ ps. The quantum dots were excited using a pulsed laser emitting at 467 nm with a pulse width ~ 200 ps and a repetition rate of 1 MHz. The PL emission was detected using a PMT after passing through a monochromator which allowed the selection of the collection wavelength. A long pass filter was placed in front of the monochromator to filter out any scattered light from the incident laser. The measurements were carried out in reflection geometry.

Steady state PL measurements were carried out to confirm the coupling of the quantum dot emission into the *high-k* modes of the metamaterial structure. For steady state PL measurements, the quantum dots were excited using a mercury lamp and the emitted light was detected using a PMT. The measurements were carried out in reflection geometry. Monchromators placed in front of the excitation source and the PMT enable the selection of excitation and emission wavelengths, respectively. Both the time resolved and steady state PL measurements were carried out using a Horiba Jobin Yvon Fluoromax system.

### Acknowledgements

Work at Queens College of CUNY was supported partially through the National Science Foundation Grant # DMR-1105392 and the Army Research Office, Grant # W911NF0710397. E.N. was partially supported Army Research Office, Grants # 50342-PH-MUR and W911NF-09-1-0539. Z.J. was partially supported by NSERC Discovery Grant # 402792. The ellipsometric measurements were carried out at the Center for Functional Nanomaterials at Brookhaven National Laboratories.

# Figure Legends

**Fig. 1** Optical topological transition (OTT): The iso-frequency curve changes from a closed surface such as an ellipsoid to a hyperboloidal open surface. This has significant consequences on the available photonic density of states which can be probed using quantum electrodynamical phenomena such as the spontaneous emission.

**Fig. 2** (a) Topological transition parameter (proportional to the local density of states) near the topological transition wavelength, the sharp transition becomes a smooth crossover as losses are increased. (b) For a practical layered realization of the metamaterial consisting of alternating layers of metal and dielectric we investigate the effect of the finite unit cell on the normalized lifetime near the transition. We see that the lifetime in the multilayer structure shows the same crossover as the unit cell size is increased (deviation from effective medium theory).

**Fig.3.** (a) Schematic of the metamaterial and (b) Effective dielectric constants (real parts) of the structure determined using effective medium theory. The transition from elliptical to hyperbolic dispersion occurs at 621 nm.

**Fig.4.** (a) Time resolved photoluminescence data from quantum dots deposited on the metamaterial, control sample and glass substrate at 605 nm, 621 nm and 635 nm clearly showing the reduction in overall lifetime over 30 nm spectral bandwidth when the emitter is placed on the metamaterial. (b) Lifetime of the quantum dots as a function of wavelength on metamaterial, control sample and glass substrate. Quantum dots on the control sample and glass shows an increase in the lifetime with wavelength, due to the size distribution of quantum dots and the weak dependence of the oscillator strength on the energy. While on the metamaterial there is almost no change in lifetime due to the presence of the *high-k* states.

**Fig.5.** The lifetime on the metamaterial normalized w.r.t the control sample which elucidates the opposing dispersion in the lifetime caused by the metamaterial states in the hyperbolic regime. The effects of quantum dot size inhomogeneity, and plasmonic reduction in a unit cell are accounted for by normalizing the lifetime w.r.t the control sample. The signature of the optical topological transition is the decreasing trend in the lifetime as compared to the glass and control sample.

# Figures

## Figure 1

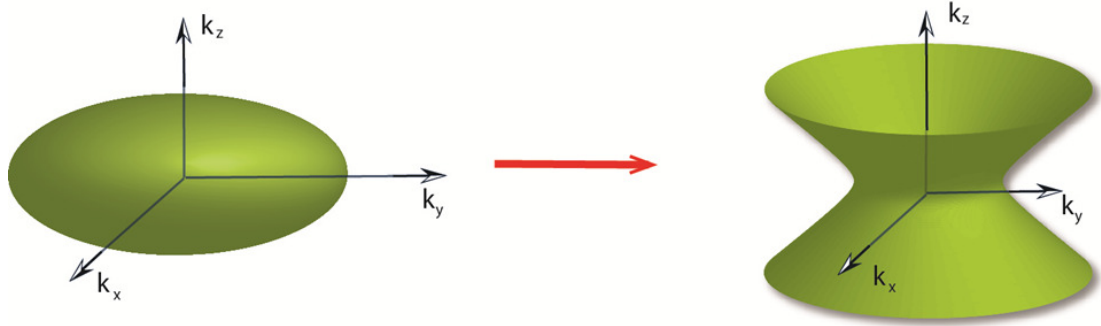

**Fig. 1** Optical topological transition (OTT): The iso-frequency curve changes from a closed surface such as an ellipsoid to a hyperboloidal open surface. This has significant consequences on the available photonic density of states which can be probed using quantum electrodynamical phenomena such as the spontaneous emission.

## Figure 2

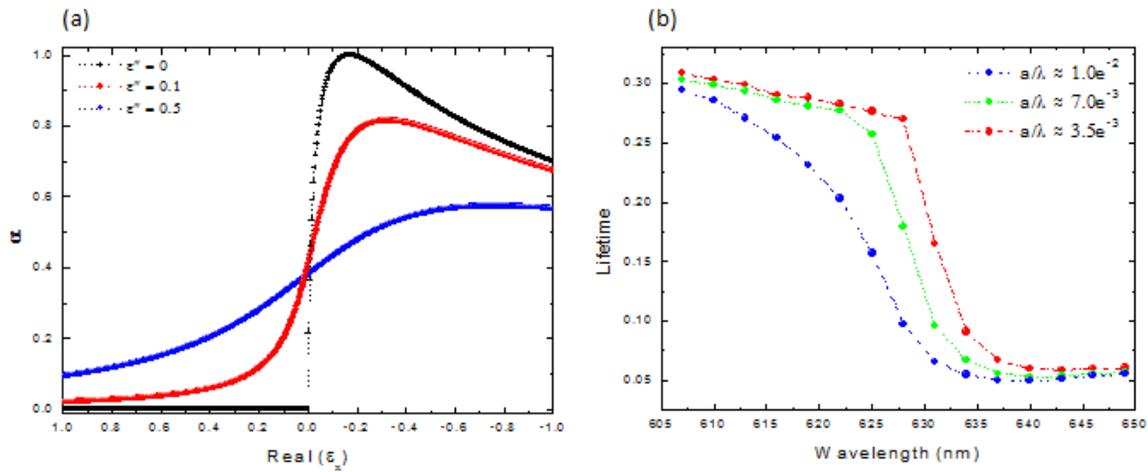

**Fig. 2** (a) Topological transition parameter (proportional to the local density of states) near the topological transition wavelength, the sharp transition becomes a smooth crossover as losses are increased. (b) For a practical layered realization of the metamaterial consisting of alternating layers of metal and dielectric we investigate the effect of the finite unit cell on the normalized lifetime near the transition. We see that the lifetime in the multilayer structure shows the same crossover as the unit cell size is increased (deviation from effective medium theory).

**Figure 3**

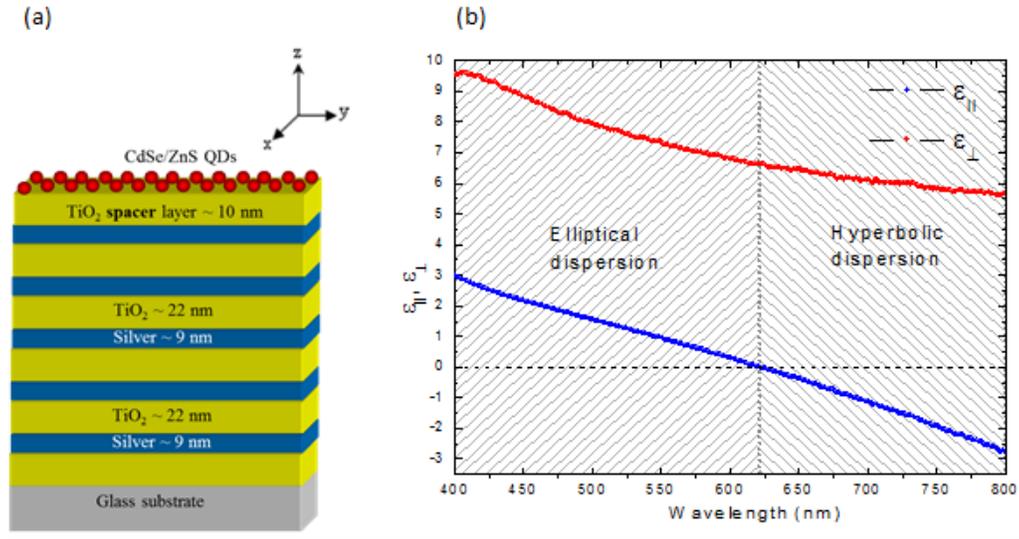

**Fig.3.** (a) Schematic of the metamaterial and (b) Effective dielectric constants (real parts) of the structure determined using effective medium theory. The transition from elliptical to hyperbolic dispersion occurs at 621 nm.

**Figure 4**

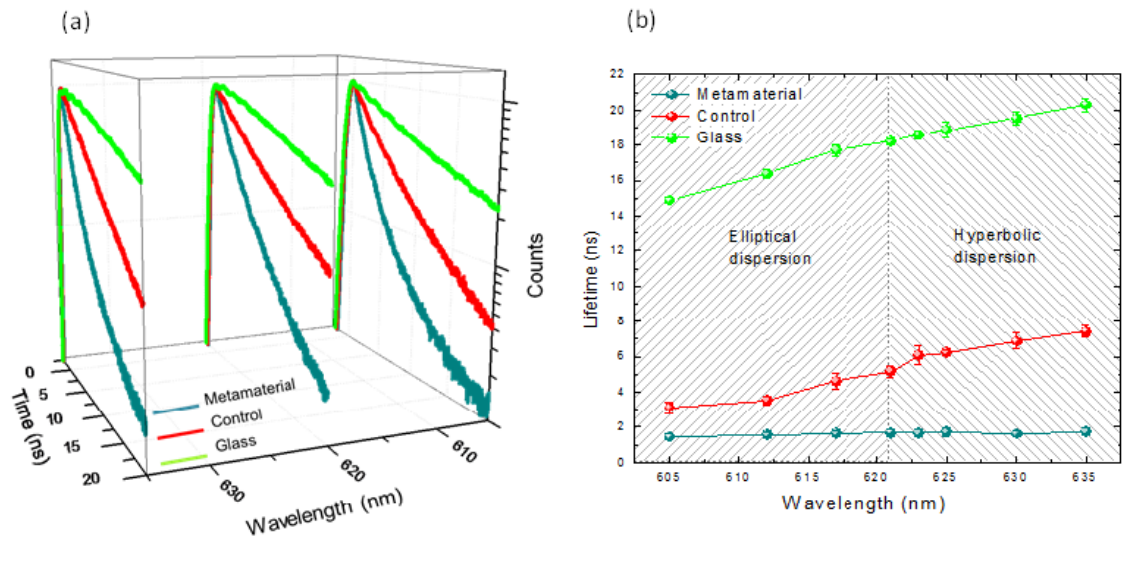

**Fig.4.** (a) Time resolved photoluminescence data from quantum dots deposited on the metamaterial, control sample and glass substrate at 605 nm, 621 nm and 635 nm clearly showing the reduction in overall lifetime over 30 nm spectral bandwidth when the emitter is placed on the metamaterial. (b) Lifetime of the quantum dots as a function of wavelength on metamaterial, control sample and glass substrate. Quantum dots on the control sample and glass shows an increase in the lifetime with wavelength, due to the size distribution of quantum dots and the weak dependence of the oscillator strength on the energy. While on the metamaterial there is almost no change in lifetime due to the presence of the *high-k* states.

**Figure 5**

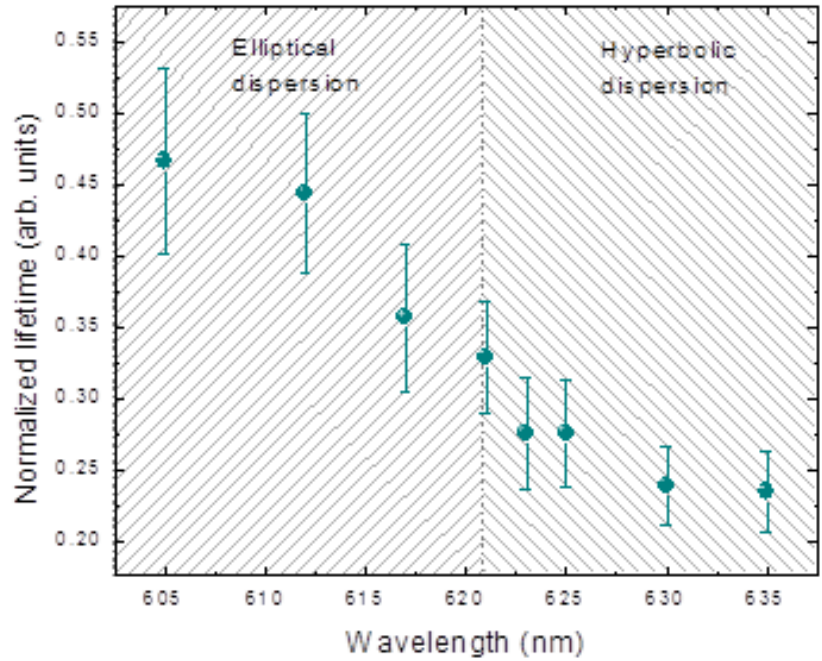

**Fig.5.** The lifetime on the metamaterial normalized w.r.t the control sample which elucidates the opposing dispersion in the lifetime caused by the metamaterial states in the hyperbolic regime. The effects of quantum dot size inhomogeneity, and plasmonic reduction in a unit cell are accounted for by normalizing the lifetime w.r.t the control sample. The signature of the optical topological transition is the decreasing trend in the lifetime as compared to the glass and control sample.